\documentclass[aps,prl,twocolumn,showpacs,superscriptaddress,groupedaddress,nofootinbib]{revtex4}  
\usepackage{bm}
\usepackage{mathrsfs}
\usepackage{amsmath}
\usepackage{amssymb}
\usepackage{graphicx}
\usepackage{amsfonts}
\usepackage{cancel}
\usepackage{amsthm}
\usepackage{color}
\usepackage{dcolumn}
\usepackage{txfonts}
\usepackage{subfigure}
\begin{document}


\title{Berry phases of quantum trajectories in semiconductors under strong terahertz fields}

\author{ Fan Yang}
\affiliation{Department of Physics, The Chinese University of Hong Kong, Shatin, N.T., Hong Kong, China}

\author{ Ren-Bao Liu }
\email{rbliu@phy.cuhk.edu.hk}
\affiliation{Department of Physics, The Chinese University of Hong Kong, Shatin, N.T., Hong Kong, China}
\affiliation{Institute of Physics, The Chinese University of Hong Kong, Shatin, N.T., Hong Kong, China}


\date{\today}

\begin{abstract}

Quantum evolution of particles under strong fields can be essentially captured by a small number of quantum trajectories that satisfy the stationary phase condition in the Dirac-Feynmann path integrals. The quantum trajectories are the key concept to understand extreme nonlinear optical phenomena, such as high-order harmonic generation (HHG), above-threshold ionization (ATI), and high-order terahertz sideband generation (HSG). While HHG and ATI have been mostly studied in atoms and molecules, the HSG in semiconductors can have interesting effects due to possible nontrivial ``vacuum" states of band materials. We find that in a semiconductor with non-vanishing Berry curvature in its energy bands, the cyclic quantum trajectories of an electron-hole pair under a strong terahertz field can accumulate Berry phases. Taking monolayer MoS$_2$ as a model system, we show that the Berry phases appear as the Faraday rotation angles of the pulse emission from the material under short-pulse excitation. This finding reveals an interesting transport effect in the extreme nonlinear optics regime.

\end{abstract}

\maketitle

After excitation by a weak near-resonant laser in semiconductors, an electron and a hole can be created and driven into large amplitude oscillations by an intense low-frequency ac electric field such as from a terahertz (THz) laser. The recollisions between the electron and hole during the oscillations will generate high-order sidebands relative to the excitation frequency~\cite{HSG_RBL,Zaks_Liu}. This high-order THz sideband generation (HSG) has potential electro-optical applications such as wide-band optical multiplexers, terabit/sec optical communications, and optical pulses with ultra-high repetition rate~\cite{HSG_RBL,Zaks_Liu}. The HSG is analogous to the high-order harmonic generation (HHG) in atoms and molecules~\cite{HHG_Krause,HHG_Corkum,HHG_QT}. Both HSG and HHG spectra are characterized by a wide-band plateau with a sharp cut-off. The HSG spectrum has been well understood using the quantum trajectory theory developed in HHG, in which the quantum evolution of particles under a strong field is described by a few paths that satisfy the stationary phase condition (i.e., the saddle points) in the formalism of Dirac-Feynmann path integrals~\cite{HSG_RBL,HHG_QT}. The HSG or HHG cutoff is determined by the maximum energy the particles in the quantum trajectories can acquire from the driving field~\cite{HSG_RBL,HHG_QT}.

A fundamental difference between HHG in atomic systems and HSG in semiconductors is that the ``vacuum" state of a
semiconductor can have non-trivial structures (such as in topological insulators~\cite{SCZhang,Kane}).
When the eletron (or hole) in a spin-orbit-coupled semiconductor is accelerated by an ac electric field ${\mathbf F}\left(t\right)$, not only does the quasi-momentum evolve according to the semiclassical equation $\dot {{\mathbf k}}=-e{\mathbf F}\left(t\right)$~\cite{Solid_AM}, but also its Bloch wavefunction (the direction of the spin) is changed. Thus the evolution driven by the electric field leads to a geometric phase in addition to the dynamical one, which is the famous Berry phase in a cyclic evolution
\cite{Berry}. This geometric phase is of fundamental importance for a gauge-invariant description of the nonlinear optics in insulators~\cite{Sipe2nd,spin_current,Sipe}.

The Berry phase effect is clearly seen in the representation of the polarization operator in the basis of Bloch states, which is found by Blount~\cite{Blount}
\begin{equation}
\left\langle {\psi _{n,{\mathbf k}} } \right|{\mathbf r}\left| {\psi _{m,{\mathbf k}'} } \right\rangle  = i\left[ {\delta_{nm} \nabla _{\mathbf k}  + \left\langle {u_{n,{\mathbf k}} } \right|\nabla _{\mathbf k} \left| {u _{m,{\mathbf k}} } \right\rangle } \right]\delta \left( {{\mathbf k} - {\mathbf k}'} \right),
\end{equation}
where the first term gives change of the quasi-momentum and the second term is the so-called Berry connection or
Berry vector potential. Through the polarization operator, the Berry phase (or, more intrinsically, the Berry curvature) appears naturally in various optical effects in condensed matter systems as revealed by some recent works. For example, the presence of the Berry phase effect was noticed in the optical birefringence effects of a pure spin current~\cite{spin_current1st} or the second-order non-linear spectroscopy of spin currents~\cite{spin_current}. Also the photogalvanic effect in topological insulator surfaces can depend on the Berry curvature~\cite{CPGE}. The interaction between an intense THz laser and semiconductors~\cite{Zaks_Liu} provides a new opportunity to explore the Berry phase effect in the regime of extreme nonlinear optics.

In this Letter, we show that the optical response of a semiconductor under an intense THz field explicitly includes the Berry phase. We analyze the effect using the quantum trajectory theory and apply the theory to monolayer $\rm{MoS}_2$ as a model system. In the time-domain response, we find that the Faraday rotation angle of the emission delayed by integer multiples of the THz laser period is given by the Berry phase of a specific trajectory. The quantum trajectory approximation is verified by numerical simulations.


Let us consider a general semiconductor under a strong THz field ${\mathbf F}(t)$, which enters into the Hamiltonian through a uniform electromagnetism vector potential: ${\mathbf p} \to {\mathbf p} + e{\mathbf A}\left( t \right)$, with ${\mathbf F}=  - \partial {\mathbf A}/\partial t$. Because ${\mathbf A}$ preserves the translational symmetry, Bloch's theorem still applies and we write the Hamiltonian in the ${\mathbf k}$-space representation $H \left(\tilde {{\mathbf k}}\left( t \right)\right)$~\cite{QNiu2},
with $\tilde {{\mathbf k}}\left( t \right) = {\mathbf k} + e{\mathbf A}\left( t \right)$. The instantaneous Bloch states of $H \left(\tilde {{\mathbf k}}\left( t \right)\right)$ are obtained from the original Bloch states by simply changing ${\mathbf k}$ to $\tilde{{\mathbf k}}$,
\begin{equation}
H\left( {\tilde {{\mathbf k}}\left( t \right)} \right)\left| { \pm ,\mu ,\tilde {{\mathbf k}}\left( t \right)} \right\rangle  = E^{\pm}_{ \tilde {{\mathbf k}}\left( t \right)}\left| { \pm ,\mu ,\tilde {{\mathbf k}}\left( t \right)} \right\rangle,
\end{equation}
where $+$ and $-$ are the indices of the conduction and valence bands, respectively, and $\mu$ is the spin index introduced to indicate possible band degeneracy of the system.

Now let us calculate the linear response of this system to a near infrared (NIR) laser that creates electron-hole pairs at the band edge with interaction Hamiltonian $\hat H_{\text{NIR}}=-\hat{{\mathbf P}}\cdot {\mathbf E}_{\text{NIR}}e^{-i\Omega t}+\text{h.c.}$, where the interband polarization operator $\hat{{\mathbf P}}$ in the interaction picture is
\begin{equation}
\hat {{\mathbf P}}\left(t\right) = \int {d{\mathbf k}} \hat e_{\mu, {\mathbf k}}^\dag \hat h_{\nu, -{\mathbf k}}^\dag {\mathscr{D}}_{\mu \nu,{\mathbf k}}\left(t\right).
\end{equation}
Here $\hat e$ and $\hat h$ are electron and hole operators, respectively, and $\mathscr{D}_{\mu \nu ,{\mathbf k}}\left(t\right)  =  - ie \left\langle {\psi_{+,\mu,{\mathbf k}}\left(t\right)} \right|\nabla _{{\mathbf k}} \left| {\psi_{-,\nu,{\mathbf k}}\left(t\right)} \right\rangle$ is the interband dipole moment, with $| {\psi_{\pm,\mu,{\mathbf k}}\left(t\right)} \rangle$ denoting the adiabatic evolution of the instantaneous Bloch states under the driving of the THz field
\begin{equation}
\left| {\psi_{\pm,\mu,{\mathbf k}}\left(t\right)} \right\rangle  = \left| {\pm, \alpha  ,\tilde {{\mathbf k}}\left( t \right)} \right\rangle \left[ {\hat T e^{ { -i\int_{ - \infty }^t E_{\tilde{{\mathbf k}} \left(\tau\right)}^{\pm}d\tau + i \int_{ - \infty }^t {\mathscr{A}_{\tilde {{\mathbf k}}\left( \tau \right)}^ {\pm} \cdot d\tilde {{\mathbf k}}\left( \tau \right)} } }} \right]_{\alpha \mu},
\end{equation}
where $\hat T$ is the time-ordering operator, the Berry connection is defined as $\left({\mathscr{A}}_{\tilde {{\mathbf k}}}^{\pm}\right)_{\mu \nu} = i\left\langle {\pm ,\mu , \tilde {{\mathbf k}}} \right|\nabla _{{\mathbf k}} \left| {\pm,\nu , \tilde {{\mathbf k}}} \right\rangle$, and summation of repeated dummy indices is assumed. In general, the Berry connection can be non-Abelian. Assuming the initial state is the vacuum state $\left|G\right\rangle$ with empty conduction bands and filled valence bands, we obtain the linear response to the NIR optical field as
\begin{align}
&\notag \left\langle {{\mathbf P}\left( t \right)} \right\rangle  = \frac{{ - i}}{V}\int_{ - \infty }^t {dt'} \left\langle G \right|\hat{{\mathbf P}}\left( t \right)\hat H_{\text{NIR}} \left( {t'} \right)\left| G \right\rangle \\
&\notag = i\int_{ - \infty }^t {dt'} \int {\frac{{d {\mathbf k}}}{{\left( {2\pi } \right)^d }}} e^{  - i\int_{t'}^t {\varepsilon _{\tilde {{\mathbf k}}\left( \tau \right)} d\tau}  - i\Omega t' }{{\mathbf d}}_{\nu \mu ,\tilde {{\mathbf k}}\left( t \right)}^\dag \left[ \hat Te^{ i \int_{t'}^t {{\mathscr{A}}_{\tilde{{\mathbf k}}\left( \tau \right)}^ + \cdot d\tilde {{\mathbf k}}\left( \tau \right)} } \right]_{\mu \mu '} \\
&\quad {{\mathbf d}}_{\mu '\nu ',\tilde {{\mathbf k}}\left( {t'} \right)}  \cdot { {\mathbf E}}_{\text{NIR}}\left[ \hat T e^{  i \int_{ t' }^t { {\mathscr{A}}_{\tilde{{\mathbf k}}\left( \tau \right)}^ -  \cdot d\tilde {{\mathbf k}}\left( \tau \right)} } \right]_{\nu '\nu }^\dag , \label{NAbelQT}
\end{align}
where $d$ is the dimension of the system, $\varepsilon _{\tilde {{\mathbf k}}}  = E^{+}_{\tilde {{\mathbf k}}}  - E^{-}_{\tilde {{\mathbf k}}}$ is the energy of the electron-hole pair and ${{\mathbf d}}_{\nu \mu,\tilde {{\mathbf k}} } = - ie \left \langle +,\nu,\tilde{{\mathbf k}}\right|\nabla _{{\mathbf k}} \left| -,\mu,\tilde{{\mathbf k}} \right\rangle$ is the instantaneous dipole moment. The Berry phase enters Eq.~(\ref{NAbelQT}) due to the requirement of the gauge invariance of the physical result under the local gauge transformation $\left| {\pm , \mu , {\mathbf k}} \right\rangle  \to \left| {\pm , \alpha , {\mathbf k}} \right\rangle U^{\pm}_{\alpha \mu}  \left( {\mathbf k} \right)$, which introduces the gauge freedom of ${{\mathbf d}}_{\nu \mu,\tilde {{\mathbf k}} }$. For the case without band degeneracy, the Berry phase becomes Abelian, and the response is reduced to
\begin{align}
\left\langle {{\mathbf P}\left( t \right)} \right\rangle = \frac{{i}}{{\left( {2\pi } \right)^d }}\int_{ - \infty }^t {dt'} \int {d {\mathbf k}} {{\mathbf d}}_{\tilde {{\mathbf k}} \left( t \right)}^* {{\mathbf d}}_{\tilde {{\mathbf k}} \left( {t'} \right)}  \cdot {{\mathbf E}}_{\text{NIR}} \notag \\
e^{ - i\int_{t'}^t {\varepsilon _{\tilde {{\mathbf k}}\left( \tau \right)} d\tau + i \int_{t'}^t {{{\mathscr {A}}_{\tilde {{\mathbf k}}  \left( \tau \right)}  } \cdot d\tilde {{\mathbf k}} \left(\tau\right)}  - i\Omega t'} } , \ \ \quad\qquad \label{AbelQT}
\end{align}
where ${\mathscr{A}}_{{\mathbf k}}={\mathscr{A}}^+_{{\mathbf k}}- {\mathscr{A}}^-_{{\mathbf k}}$ is the combined Berry connection of the electron-hole pair. It is interesting to note that the Berry term $\mathscr{A}_{\tilde {{\mathbf k}} } \cdot {\dot {\tilde {{\mathbf k}}}}$ in the phase factor can be regarded as a generalization of the interaction energy ${\mathbf A} \cdot \dot{{\mathbf r}}$ in the electromagnetism theory in real space to the momentum space~\cite{QNiu2}.

Now let us focus on the geometric phase part of Eq.~(\ref{AbelQT}). Without loss of generality, we consider an elliptically polarized THz field in the $x$-$y$ plane
\begin{equation}\label{Ft}
{\mathbf F}\left( t \right) = F \left( \cos\left(\theta\right)\cos \left(\omega t\right),\sin\left(\theta\right)\sin \left(\omega t\right), 0 \right).
\end{equation}
Then the electron-hole pair goes along an elliptical path in the ${\mathbf k}$-space under the driving of the THz field:
\begin{equation}
\tilde {{\mathbf k}}\left( t \right) = \left( {k_x  - k_0\cos\theta \sin \left(\omega t\right),k_y  + k_0\sin\theta \cos \left(\omega t\right)},k_z \right), \label{path}
\end{equation}
where $k_0  = {eF}/\omega$. After an integer multiple periods of the THz field $t-t'=nT=2n\pi/\omega$,
the electron-hole pair completes a closed path in the ${\mathbf k}$-space. During this cyclic evolution, the geometric phase acquired by the electron-hole pair equals the Berry curvature times the area $S\left(\theta,k_0\right)=n\pi k_0^2 \sin\left(2\theta\right)/2$ enclosed by the path, which is further related to the polarization and strength of the THz light.

To measure the Berry phase effect of quantum trajectories, we can apply a short NIR laser pulse to the system at time $t'=t_0$, with the width of the pulse much smaller than $T$. Thus the pulse can be approximated by a $\delta$-pulse ${{\mathbf E}}_{\text{NIR}}={\mathbf E}\delta\left(t-t_0\right)$. Then response of the system at $t_n=t_0+nT$ explicitly contains the Berry phase of a closed path: $\phi_B^{\left(n\right)}\left({\mathbf k}\right)=\int_{t_0}^{t_n} { {\mathscr{A}}_{\tilde {{\mathbf k}}\left( \tau \right)}  \cdot d\tilde {{\mathbf k}}\left( \tau \right)}$. In order to separate the Berry phase from the dynamical phase, we can introduce some specific interference that singles out the geometric phase part. We note that under the time-reversal transformation, the direction of the path in ${\mathbf k}$-space is reversed and the Berry phase becomes opposite, while the dynamical phase is unchanged. It leads us to consider the solid state systems that preserve the time-reversal and inversion symmetry and have nontrivial Berry phases, such as the topological insulators~\cite{SCZhang,Kane}, monolayer $\rm{MoS}_2$ and other group-VI dichalcogenides~\cite{MOS2_Mak1,MOS2,MOS2_Zeng,MOS2_Mak,MOS2_Cao} and bilayer graphene~\cite{Graphene_RMP,BilayerG}. We denote one state of the Kramers pair as the pseudospin state $\Uparrow$ and the other as $\Downarrow$. From the time-reversal and inversion symmetry, we obtain the following relations ${{\mathbf d}}_{ \Uparrow  \Uparrow ,{\mathbf k}}  = {{\mathbf d}}_{ \Downarrow  \Downarrow , - {\mathbf k}}^*  =  - {{\mathbf d}}_{ \Downarrow  \Downarrow ,{\mathbf k}}^*:={{\mathbf d}}_{{\mathbf k}}$, $\varepsilon _{ \Uparrow , {\mathbf k}}  = \varepsilon _{ \Downarrow , - {\mathbf k}}: = \varepsilon _{{\mathbf k}}$ and $\left({\mathscr{A}}_{{\mathbf k}}\right) _{\Uparrow  \Uparrow}  = \left({\mathscr{A}}_{ - {\mathbf k}}\right)_{\Downarrow  \Downarrow}^*:={\mathscr{A}}_{{\mathbf k}}$. Thus we get the key formula
\begin{equation}
\phi_B^{\left(n\right)}\left({\mathbf k}\right)=\phi^{\left(n\right)}_{B,\Uparrow\Uparrow}\left({\mathbf k}\right)
=-\phi^{\left(n\right)}_{B,\Downarrow\Downarrow}\left({\mathbf k}\right)
=\int_{t_0}^{t_n} {{\mathscr{A}}_{\tilde {{\mathbf k}}}  \cdot d\tilde {{\mathbf k}}},
\end{equation}
i.e. the Berry phase of the two time-reversal related paths are opposite to each other. With these considerations, the response at $t_n$ is simplified as
\begin{align}
\left\langle {{\mathbf P}\left( t_n \right)} \right\rangle = & i\int {\frac{d{{\mathbf k}}}{{\left( {2\pi } \right)^d }}} e^{ { - i\int_{t_0}^{t_n} {\varepsilon _{\tilde {{\mathbf k}}\left( \tau \right)} d\tau}  +i \phi_B^{\left(n\right)}\left({\mathbf k}\right)  } } {{\mathbf d}}_{\tilde {{\mathbf k}}\left( t_0 \right)}^* {{\mathbf d}}_{\tilde {{\mathbf k}}\left( t_0 \right)}  \cdot { {\mathbf E}} \ \notag\\
+ & i\int {\frac{d{{\mathbf k}}}{{\left( {2\pi } \right)^d }}} e^{ { - i\int_{t_0}^{t_n} {\varepsilon _{\tilde {{\mathbf k}}\left( \tau \right)} d\tau}  -i \phi_B^{\left(n\right)}\left({\mathbf k}\right)  } } {{\mathbf d}}_{\tilde {{\mathbf k}}\left( t_0 \right)}{{\mathbf d}}^*_{\tilde {{\mathbf k}}\left( t_0 \right)}  \cdot { {\mathbf E}}. \label{RespN}
\end{align}
Here $\left\langle {{\mathbf P}\left( t_n \right)} \right\rangle$ is given by the interference between two kinds of responses with opposite Berry phases.

\begin{figure}
\begin{center}
\includegraphics[width=\columnwidth]{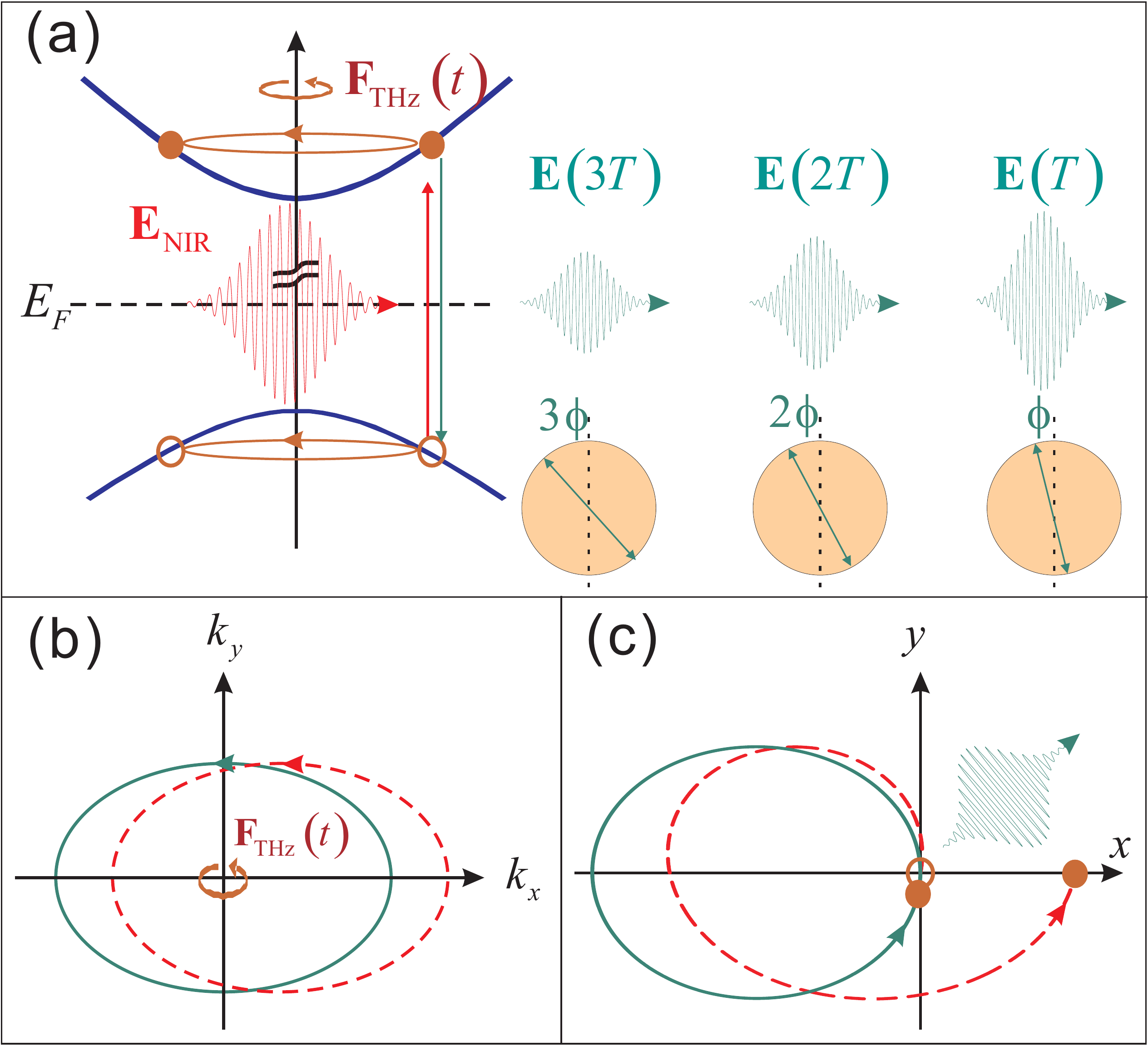}
\end{center}
\caption{(color online). Schematics of optical response to a short NIR pulse and quantum trajectories in semiconductors under strong THz fields. (a) An electron-hole pair is excited by an NIR pulse and then driven along quantum trajectories by an elliptically polarized THz field. At $t_n-t_0=nT$, the electron-hole pair completes an elliptical loop in ${\mathbf k}$-space (see (b)) and obtains a nonzero Berry phase. The polarization of optical emission at $t_n$ is rotated by an angle $n\phi$ that is equal to the Berry phase $\phi^{\left(n\right)}_B$. (b) Paths of the electron-hole pairs in ${\mathbf k}$-space, where the green solid curve is for the electron-hole pair satisfying the stationary phase condition in Eq.~(\ref{saddle}). The red dashed one does not satisfy Eq.~(\ref{saddle}). (c) The green solid curve gives the quantum trajectory of an electron-hole pair that recombines at $t_1$. The red dashed curve corresponds to the red dashed path in (b), in which the electron-hole pair does not make a close path in real space.}
\label{Schem}
\end{figure}

Equation~(\ref{RespN}) can be studied using the quantum trajectory theory, i.e. the stationary phase formalism~\cite{HSG_RBL,HHG_QT}. In the path integral, the electron-hole pairs move along all possible trajectories when driven by the THz field, with the phase given by the action $S_{\pm}\left({\mathbf k}\right)= \int_{t_0}^{t_n} \left(\varepsilon _{\tilde {{\mathbf k}}\left( \tau \right)}\pm{\mathscr{A}}_{\tilde {{\mathbf k}}}  \cdot e{{\mathbf F}}\left(\tau\right) \right) d\tau$. As the THz field is strong, the motion amplitude of the electron-hole pair is much larger than the quantum fluctuation. Thus the response is dominated by the stationary phase points of the actions
\begin{equation}
\nabla_{{\mathbf k}}S_{\pm}\left({\mathbf k}\right)= \int_{t_0}^{t_n} \left(\nabla_{{\mathbf k}}\varepsilon _{\tilde {{\mathbf k}}\left( \tau \right)} \pm \nabla_{{\mathbf k}}{\mathscr{A}}_{\tilde {{\mathbf k}}}  \cdot e{{\mathbf F}}\left(\tau\right) \right) d\tau=0 , \label{saddle}
\end{equation}
plus the Gaussian quantum fluctuation around them.  $\nabla_{{\mathbf k}}{\varepsilon}_{\tilde{{\mathbf k}}}={{\mathbf v}}_{\tilde{{\mathbf k}}}$ is the semiclassical velocity of the electron-hole pair. The Berry connection term gives a gauge dependent motion $\nabla _{{\mathbf k}} \left({\mathscr{A}}_{\tilde {{\mathbf k}}}  \cdot e{{\mathbf F}}\right)= \nabla_{{\mathbf k}} \left({\mathscr{A}}_{\tilde {{\mathbf k}}} \cdot e{{\mathbf F}}\right) - \left(e{{\mathbf F}}\cdot \nabla_{{\mathbf k}}\right) {\mathscr{A}}_{\tilde {{\mathbf k}}}  - \frac{d}{{d\tau }}{\mathscr{A}}_{\tilde {{\mathbf k}}}$. Because we are considering the cyclic evolution along a closed loop, the last gauge dependent term vanishes and the first two terms gives a gauge-invariant physical quantity $\Omega _{\tilde k_i \tilde k_j } eF_j {{\mathbf e}}_i$, with $\Omega _{\tilde k_i \tilde k_j }
=\partial_{k_i}{\mathscr{A}}_{\tilde {k_j}}-\partial_{k_j}{\mathscr{A}}_{\tilde {k_i}}$ being the Berry curvature. We note that this is the well-known anomalous velocity that is responsible for various Hall effects~\cite{Chang_Niu1995,Sundaram_Niu,QNiu2}. The stationary phase condition in Eq.~(\ref{saddle}) therefore means the return of the electron to the hole after $nT$ under the acceleration by the THz field. The Berry phase  $\phi_B^{\left(n\right)}\left({\mathbf k}\right)$ in the actions is generally a slowly-varying function of ${\mathbf k}$ and much smaller than the dynamical phase factor. Therefore the stationary phase points are determined by $\int_{t_0}^{t_n} {{{\mathbf v}}_{\tilde {{\mathbf k}}\left( \tau \right)} d\tau}=0$, which has a simple solution ${{\mathbf v}}_{{\mathbf k}}=0$ if the effective mass model is used. This means that the response is dominated by the trajectories of the electron-hole pairs whose paths in the ${\mathbf k}$-space are centered at the extreme points of the energy band (see Fig. \ref{Schem}). Thus Eq.~(\ref{RespN}) is approximated by
\begin{align}
 & \notag \left\langle {{\mathbf P}\left( t_n \right)} \right\rangle \\
\approx & \cos\left(\phi_B^{\left(n\right)}\right)_{{{\mathbf v}}_{{\mathbf k}}=0}\int {\frac{2id{{\mathbf k}}}{{\left( {2\pi } \right)^d }}} e^{ { - i\int_{t_0}^{t_n} {\varepsilon _{\tilde {{\mathbf k}}\left( \tau \right)} d\tau}  } } \Re\left[{{\mathbf d}}_{\tilde {{\mathbf k}}\left( t_0 \right)}^* {{\mathbf d}}_{\tilde {{\mathbf k}}\left( t_0 \right)} \cdot { {\mathbf E}}\right]\notag \\
\ - &\sin\left(\phi_B^{\left(n\right)}\right)_{{{\mathbf v}}_{{\mathbf k}}=0}\int {\frac{2id{{\mathbf k}}}{{\left( {2\pi } \right)^d }}} e^{ { - i\int_{t_0}^{t_n} {\varepsilon _{\tilde {{\mathbf k}}\left( \tau \right)} d\tau}  } } \Im\left[{{\mathbf d}}_{\tilde {{\mathbf k}}\left( t_0 \right)}^* {{\mathbf d}}_{\tilde {{\mathbf k}}\left( t_0 \right)}\cdot { {\mathbf E}}\right]  .
\label{RespQT}
\end{align}

To be specific, from now on we consider the monolayer $\rm{MoS}_2$ as the model system. This material has interesting spin-valley coupling and has potential applications for novel spin- and valley-based information processing~\cite{MOS2_Mak1,MOS2,MOS2_Zeng,MOS2_Mak,MOS2_Cao}. The effective Hamiltonian describing the Bloch states at the band edges is given by~\cite{MOS2}
\begin{equation}
H\left( {\mathbf k} \right) = A\left( {\tau k_x \sigma _x  + k_y \sigma _y } \right) + M\sigma _z,
\end{equation}
where $2M=1.59 \ {\rm{eV}}$ is the band gap, $A=3.51 \ {\rm{eV}} \cdot \AA$ and $\tau=\pm 1$ is the index of the $\pm K$ valley. The energy spectrum is $\varepsilon_{{\mathbf k}}=2\sqrt{M^2+A^2 k^2}$ and the stationary phase point is ${\mathbf k}=0$. The two valleys are related by time-reversal transformation and we denote the state at the $-K$ valley as the pseudospin state $\Uparrow$. We choose the gauge such that the Berry connection ${\mathscr{A}}_{{\mathbf k}}  = \frac{\varepsilon_{{\mathbf k}}-2M}{\varepsilon_{{\mathbf k}}k^2}\left( {k_x {{\mathbf e}}_y  - k_y {{\mathbf e}}_x } \right)$ and the dipole moment ${{\mathbf d}}_{{\mathbf k}} \approx d_{cv,{\mathbf k}}\left( {{\mathbf e}}_x  +i{{\mathbf e}}_y \right)$ for small ${\mathbf k}$ in the $-K$ valley. For the sake of simplicity we have not included the Coulomb interaction between electrons and holes. This is justified for the band edge excitation since the exciton binding energy (100s of meV~\cite{note}) is much greater than the THz field and therefore the exciton bound states are far off-resonant from the NIR excitation.

We assume that the NIR field is linearly polarized in the $x$-$y$ plane with ${\mathbf E}=E{\mathbf e}_{\parallel}$. The dipole moment gives
\begin{equation}
{{\mathbf d}}_{{\mathbf k}}^* {{\mathbf d}}_{{\mathbf k}} = \left| {d_{cv,{\mathbf k}} } \right|^2 \left[ \left( {{\mathbf e}}_x {{\mathbf e}}_x  + {{\mathbf e}}_y {{\mathbf e}}_y \right) + i\left( {{\mathbf e}}_x {{\mathbf e}}_y  - {{\mathbf e}}_y {{\mathbf e}}_x  \right) \right]. \label{Spe_dipole}
\end{equation}
The real part of~(\ref{Spe_dipole}) leads to the longitudinal response along ${\mathbf e}_{\parallel}$ while the imaginary part leads to the transverse response along ${\mathbf e}_{\bot}$ (which is related to the Faraday rotation of the emission) with ${{\mathbf e}}_\parallel   \times {{\mathbf e}}_ \bot   = {{\mathbf e}}_z$. Thus the longitudinal response of Eq.~(\ref{RespQT}) is
\begin{equation}
\left\langle {{\mathbf P}\left( t_n \right)} \right\rangle_{\parallel}= \cos\left(\phi_B^{\left(n\right)}\right)_{{\mathbf k}=0} \int {\frac{2id{{\mathbf k}}}{{\left( {2\pi } \right)^d }}} e^{ { - i\int_{t_0}^{t_n} {\varepsilon _{\tilde {{\mathbf k}}\left( \tau \right)} d\tau}  } } \left|d_{cv,\tilde {{\mathbf k}}\left( t_0 \right)}\right|^2 E ,
\end{equation}
and the transverse response is
\begin{equation}
\left\langle {{\mathbf P}\left( t_n \right)} \right\rangle_{\bot}= \sin\left(\phi_B^{\left(n\right)}\right)_{{\mathbf k}=0} \int {\frac{2id{{\mathbf k}}}{{\left( {2\pi } \right)^d }}} e^{ { - i\int_{t_0}^{t_n} {\varepsilon _{\tilde {{\mathbf k}}\left( \tau \right)} d\tau}  } } \left|d_{cv,\tilde {{\mathbf k}}\left( t_0 \right)}\right|^2 E .
\end{equation}
The Faraday rotation angle of the optical emission is exactly given by the Berry phase
\begin{equation}
{\phi}_{r}^{\left(n\right)}\left(\theta,k_0\right) = \phi_B^{\left(n\right)}\left({\mathbf k}=0\right) \approx \frac{{n\pi A^2 k_0^2 }}{{2M^2 }}\sin \left( {2\theta } \right), \label{Rotangle}
\end{equation}
where $A^2/M^2$ is the Berry curvature for small ${\mathbf k}$~\cite{MOS2}.

\begin{figure}[b]
\begin{center}
\includegraphics[width=\columnwidth]{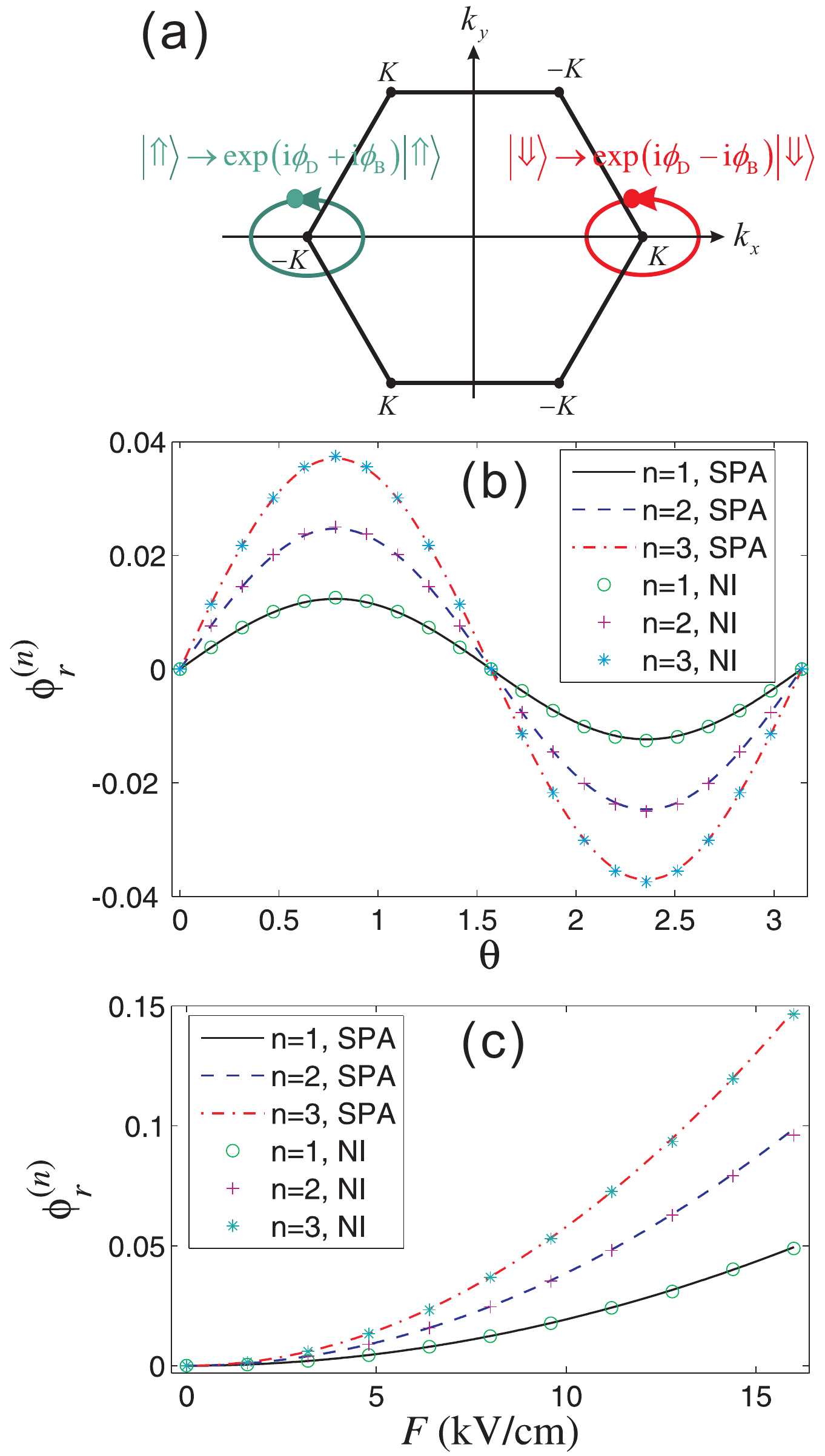}
\end{center}
\caption{(color online). Faraday rotation of the optical emission of monolayer $\rm{MoS}_2$ under a strong THz field. (a) Schematic for opposite Berry phases of quantum trajectories in different valleys. (b) and (c) plot Faraday rotation angle ${\phi}_{r}^{\left(n\right)}$ of the optical emission at $t=nT$ ($n=1,2,3$) as a function of the polarization ellipticity ($\theta$) and the strength ($F$) of the THz field, respectively. The symbols show the numerical integration (NI) results and the lines are the stationary phase approximation (SPA) in Eq.~(\ref{Rotangle}). In (b), the THz field strength $F=8\ {\rm{kV/cm}}$ (i.e., $k_0=0.02$). In (c), the THz field is circularly polarized (i.e., $\theta=\pi/4$).}
\label{Phir_fig}
\end{figure}

The Faraday rotation effect of an elliptically polarized THz field can be intuitively understood as illustrated in Fig.~\ref{Phir_fig}(a).
The electron-hole pair created by a linearly polarized short NIR pulse excitation is a superposition of the valley states $\left|  \Uparrow  \right\rangle  + \left|  \Downarrow  \right\rangle$. After the cyclic evolution under the THz field, the states at $\mp K$ valleys (corresponding to $\tau=\pm 1$) obtain the same dynamical phase $\phi_D$ and opposite Berry phases $\pm\phi_B$, which are the Berry curvature fluxes through the area enclosed by the quantum trajectories. Thus the final state is $e^{i\phi _D } \left( {e^{i\phi _B} \left|  \Uparrow  \right\rangle  + e^{ - i\phi _B } \left|  \Downarrow  \right\rangle } \right)$, which results in emission with linear polarization rotated by an angle $\phi_B$.

Equation~(\ref{Rotangle}) shows clearly the effect of the Berry phase on the optical response of the semiconductor in an intense THz field and provides a new method to directly measure the Berry phase of the energy bands in momentum space. Since the Faraday rotation angle is independent of the strength of the optical response, the measurement does not rely on the specific form of the energy spectrum, the value of $d_{cv,{\mathbf k}}$, or the dephasing of the electron-hole pair during the evolution. However, the result does depend on the optical selection rule of the dipole moment ${{\mathbf d}}_{{\mathbf k}}\sim \left( {{\mathbf e}}_x  +i{{\mathbf e}}_y \right)$, which is due to the rotational symmetry of the system. Thus Eq.~(\ref{Rotangle}) may be applied to other two-dimensional spin-orbit coupled semiconductors with (approximate) rotational symmetry. For a semiconductor having a different form of the dipole moment, the Faraday rotation angle of the optical emission may have a more complex relation to the Berry phase.

To verify the validity of Eq.~(\ref{Rotangle}), we compare it with the numerical results obtained from the standard numerical integration of Eq.~(\ref{AbelQT}) for a $\rm{MoS}_2$ monolayer.
The frequency of the THz field is $\omega=4 \ \text{meV}$ and the NIR pulse has the gaussian form $E{{\mathbf e}}_x \exp \left( { - i\Omega t  - t^2 /\delta t ^2 } \right)$, where $\Omega=2M$ and the width of the pulse is such that $\omega \delta t  = 0.2$ ($\ll 2\pi$). Some results calculated for different $n$, $\theta$ and field strength $F$ (i.e. $k_0$) are shown in Fig. \ref{Phir_fig}, where the lines are the stationary phase approximation and the symbols give the numerical integration results. We can see that Eq.~(\ref{Rotangle}) is a good approximation.

In the discussions above, we only consider the emissions at $t=t_0 + nT$. However, in the case of $t-t_0 \ne nT$, the trajectories of the electron-hole pairs also obtain a geometric phase. For the monolayer $\rm{MoS}_2$, the dipole moment is nearly constant for small ${\mathbf k}$ in the gauge we chose above. Then based on the same reasoning, we see that the Faraday rotation angle of the emission at any time $t$ is given by the geometric phase of the  trajectory that satisfies the stationary phase point equation $\int_{t_0}^{t} {{{\mathbf v}}_{\tilde {{\mathbf k}}\left( \tau \right)} d\tau}=0$.

In summary, we have obtained a Berry phase dependent theory of optical response in spin-orbit-coupled semiconductors under strong THz fields, where the Berry phase enters the formula as required by the gauge invariance of the optical response. This theory is investigated using the quantum trajectory theory and applied to the monolayer $\rm{MoS}_2$. The Faraday rotation angle of the optical emission is exactly equal to the Berry phase of the quantum trajectory that satisfies the stationary phase condition. This result can be generalized to semiconductors without time-reversal symmetry. Even more interesting, the theory can be applied to semiconductors with non-Abelian Berry connection such as the three-dimensional topological insulators~\cite{SCZhang,Kane}. The quantum trajectory will then be accompanied by a nontrivial (pseudo)spin rotation that is determined by the non-Abelian Berry phase.

This work is supported by Hong Kong RGC/GRF 401512 and the CUHK Focused Investments Scheme.

\newcommand{\noopsort}[1]{} \newcommand{\printfirst}[2]{#1}
  \newcommand{\singleletter}[1]{#1} \newcommand{\switchargs}[2]{#2#1}

\end{document}